\documentclass[12pt]{iopart}
\newcommand{\CPG}{Ce$_2$Ga$_{12}$Pt}
\usepackage{graphicx}  
\usepackage{amssymb}
\begin{document}

\title{Physical properties and crystal chemistry of \CPG}

\author{O Sichevych$^{1,2}$, C Krellner$^{1,3}$, Yu Prots$^1$, Yu Grin$^{1,*}$, F Steglich$^{1,+}$}

\address{$^1$ Max Planck Institute for Chemical Physics of Solids, Noethnitzer Str. 40, 01187 Dresden, Germany\\
$^2$ Department of Chemistry, National University of Forestry and Wood Technology of Ukraine, General Chuprinky Street 103, 79013 Lviv, Ukraine\\
$^3$ Cavendish Laboratory, University of Cambridge, J J Thomson Avenue, Cambridge CB3 0HE, United Kingdom\\}
\ead{$^*$grin@cpfs.mpg.de, $^+$steglich@cpfs.mpg.de}

\begin{abstract}
Single crystals of the new ternary compound \CPG\, were prepared by the self-flux technique. The crystal structure with the space group $P4/nbm$ was established from single-crystal X-ray diffraction data and presents a derivative of the LaGa$_6$Ni$_{0.6}$ prototype. 	
Magnetic susceptibility measurements show  Curie-Weiss behaviour due to local Ce$^{3+}$ moments. At high temperatures, the magnetic anisotropy is dominated by the crystal-electric-field (CEF) effect with the easy axis along the crystallographic $c$ direction. \CPG\, undergoes two antiferromagnetic phase transitions at $T_{N,1} = 7.3$\,K and $T_{N,2} = 5.5$\,K and presents several metamagnetic transitions for the magnetic field along $c$. Specific-heat measurements prove the bulk nature of these magnetic transitions and reveal a doublet CEF ground state. The $4f$ contribution to the resistivity shows a broad maximum at $T_{max}\approx 85$\,K due to Kondo scattering off the CEF ground state and excited levels.
\end{abstract}

\maketitle

\section{Introduction}
The investigation of new cerium-based intermetallic compounds is a fascinating research area owing to a large variety in their physical ground states. Phenomena like valence fluctuations, heavy fermion behaviour, unconventional superconductivity and localized magnetism are frequently observed \cite{Stewart:2001}. Recently, systems with reduced dimensionality or frustration are attracting strong interest because these features lead to an enhancement of quantum fluctuations which often result in unusual, very interesting properties. In particular, an increase of the superconducting transition temperature when going from three-dimensional to more two-dimensional compounds was predicted \cite{Monthoux:2001} and indeed observed among different classes of unconventional superconductors with magnetic fluctuations, e.g., CeCoIn$_5$ \cite{Petrovic:2001} and the iron arsenides \cite{Mazin:2009}.

Ternary cerium compounds with gallium and $4d$ transition metals have so far only scarcely been investigated \cite{Macaluso:2005}. More detailed studies on the crystal structures and physical properties were performed on compounds containing Ni, Co, Fe, Mn \cite{Grin:1989}. With Ni the existence of the gallium-rich CeGa$_6$Ni$_{0.6}$  \cite{Grin:1982}, Ce$_2$Ga$_{10}$Ni \cite{Yarmolyuk:1982}, Ce$_4$Ga$_{17}$Ni$_2$ \cite{Grin:1983}, and Ce$_3$Ga$_{15}$Ni$_2$ \cite{Grin:1984} was shown, with different stacking patterns of atomic slabs along one crystallographic direction. 

Only few physical characterization studies on ternary gallides were reported: CeGa$_6$Pd presents a broad antiferromagnetic (AFM) phase transition at $T_N=5.5$\,K with an enhanced Sommerfeld coefficient of the electronic specific heat \cite{Macaluso:2005}. It exhibits magnetic anisotropy and a pronounced metamagnetic transition when the field is applied along the $c$ direction. The related compound Ce$_2$Ga$_{12}$Pd  orders antiferromagnetically at $T_N=11$\,K and presents a similar metamagnetic transition. Further members of this compound family $RE_2$Ga$_{12}TM$ with $RE =$ La, Ce and $TM =$ Ni, Cu were also investigated \cite{Kaczorowski:2002, Cho:2008}.

Here, we report on the physical and structural properties of a new, related compound \CPG\, showing disorder in the gallium substructure. Detailed magnetic, transport and thermodynamic measurements on single crystals reveal local Ce$^{3+}$ moments which undergo two successive AFM transitions at $T_{N,1} = 7.3$\,K and \nolinebreak $T_{N,2}\nolinebreak = \nolinebreak 5.5$\,K.

\section{Experimental}
\subsection{Single crystal growth}
Single crystals of \CPG\, were obtained by using a flux growth method. Ce ingots (Ames, 99\%), Pt foil (Chempur, 99.95\%), Ga pieces (Chempur, 99.999\%) were used in a 1:1.05:20 ratio. For preparation, the reaction mixture was put inside an Al$_2$O$_3$ crucible and sealed in a Ta tube with a sieve. The tube was enclosed in a quartz ampoule and kept at 1100$^{\circ}$C for 2\,h. Subsequently, the melt was cooled down to 450$^{\circ}$C with a rate of 5\,K/h, at this point the ampoule was immediately inverted and centrifuged. The plate-like single crystals were mechanically extracted from the sieve.

\subsection{X-ray diffraction} 
Single crystal X-ray diffraction data were collected on a Rigaku AFC7 diffraction system (Mo$K_{\alpha}$ radiation, $\lambda = 0.71073$\,\AA). X-ray powder diffraction patterns of the powdered single crystals were performed on a Huber Imaging Plate Guinier Camera G670 using Cu$K_{\alpha 1}$ radiation   ($\lambda = 1.54060$\,\AA). Unit-cell parameters were refined by a least squares procedure using the peak positions extracted from powder patterns measured with LaB$_6$ as internal standard ($a = 4.15692$\,\AA). Indexing of the diffraction peaks in the powder diagrams was controlled by intensity calculations using positional parameters of the refined crystal structures. All crystallographic calculations were performed with the program package WinCSD \cite{Akselrud:1993}.

\subsection{Physical property measurements}
Magnetic measurements were performed in a commercial Quantum Design (QD) magnetic property measurement system (MPMS) with a 5\,T magnet as well as in a QD VSM SQUID equipped with a 7\,T magnet. The resistivity, $\rho(T)$, was determined down to 1.8\,K using a standard \textit{ac} four-probe geometry in a QD physical property measurement system (PPMS). The PPMS was also used to measure the specific heat, $C(T)$, with a  heat-pulse relaxation technique.

\section{Results and Discussion}
\subsection{Crystal structure}

\begin{table}
\begin{center}
\caption{\label{tab1}Data collection and structure refinement parameters for \CPG.}
\begin{tabular}{@{}ll}
\br
Composition&\CPG\\
\mr
Space group & $P4/nbm$\\
$Z$ & $2$\\
$a$, \AA $^{*}$&	$6.1004(2)$\\
$c$, \AA $^{*}$&	$15.5961(7)$\\
Unit cell volume, \AA$^3$ &	$580.41(1)$\\
Calculated density (g/cm$^3$) &	$7.420(1)$\\
Absorption coefficient (1/cm) &	$485.90$\\
Radiation and wavelength, \AA & Mo$K_{\alpha}$\, $0.71073$\\
Diffractometer	& Rigaku AFC7\\
$2\theta_{max}$ & $64.73$\\
$N(hkl)$ measured & $5158$\\ 
$N(hkl)$ symmetrically independent & 566 ($R_{int}=0.031$)\\
$N(hkl)$ used for refinement ($I > 2\sigma(I)$)  & 511\\
Mode refinement	& $F(hkl)$\\
Restrictions	& $F (hkl) > 4\sigma(F)$\\
Refined parameters & 26\\
$R(F)$	& 0.024\\
\br
\end{tabular}
\begin{tabular}{l}
$^{*}$Lattice parameters obtained from X-ray powder diffraction data.
\end{tabular}
\end{center}
\end{table}

\small
\begin{table}
\begin{center}
\caption{\label{tab2}Atomic coordinates and displacement parameters for \CPG.}
\begin{tabular}{@{}lllllllllll}
\br
At. &	Wy.  & Occ. & $ x/a $ & $y/b$ & $z/c$ &  $B_{eq}$\,$^*$ & $B_{11}$ & $B_{33}$ & $B_{12}$ & $B_{13}$\\
& site & & & & & & $=B_{22}$ & & & = -$B_{23}$ \\ 
\mr
Ce	& $4h$ & 1 & \footnotesize{$^3/_4$} & \footnotesize{$^1/_4$}  & $0.24592(4)$ & 	$0.73(1)$ & $0.64(2)$ & $0.92(2)$ & -$0.00(2)$ & 0 \\
Pt	& $2c$ & 1 & \footnotesize{$^3$/$_4$} & \footnotesize{$^1/_4$}  & $0$ &	$0.69(1)$ & $0.61(2)$ & $0.83(2)$ & $0$ & 0 \\
Ga1	& $8m$ & 1 & $0.5003(2)$ & -$x$  & $0.08808(5)$ &	$0.80(1)$ & $0.82(2)$ & $0.76(3)$ & $0.16(2)$ & $0.02(2)$ \\
Ga2	& $4g$ & 1 & \footnotesize{$^1/_4$} & \footnotesize{$^1/_4$}  & $0.18437(8)$ &	$0.70(2)$ & $0.68(3)$ & $0.72(4)$ & $0$ & $0$ \\
Ga3	& $4g$ & 1 & \footnotesize{$^1/_4$} & \footnotesize{$^1/_4$}  & $0.34179(8)$ &	$1.12(2)$ & $1.23(3)$ & $0.90(4)$ & $0$ & $0$ \\
Ga4	& $8m$ & $0.56(1)$ & $0.5863(2)$ & $x$\,-\,\footnotesize{$^1/_2$}  & $0.4299(2)$ &	$0.96(6)$ & $0.8(1)$  & $1.32(9)$  & -$0.4(1)$  & $0.2(1)$ \\
Ga5	& $8m$ & $0.39(1)$ & $0.5451(3)$ & $x$\,-\,\footnotesize{$^1/_2$}  & $0.4270(2)$ &	$0.75(8)$ & $0.7(1)$  & $0.9(1)$  & -$0.5(1)$  & $0.3(1)$ \\
Ga6	& $8m$ & $0.025(3)$ & $0.952(5)$ & -$x$  & $0.423(2)$ &	$0.7(6)$ &  &  &  &  \\
Ga7	& $2c$ & $0.034(6)$ & \footnotesize{$^3/_4$} & \footnotesize{$^1/_4$}  & \footnotesize{$^1/_2$} &	$0.8(8)$ &  &  &  &  \\
\br
\end{tabular}
\begin{tabular}{l}
$^*$\, $B_{eq}=4/3\,[B_{11}(a^*)^2a^2+...+2B_{23}(b^*)(c^*)\,b\,c \cos \alpha]$
\end{tabular}
\end{center}
\end{table}
\normalsize

\begin{figure}[t]
\begin{center}
\includegraphics[width=0.6\textwidth]{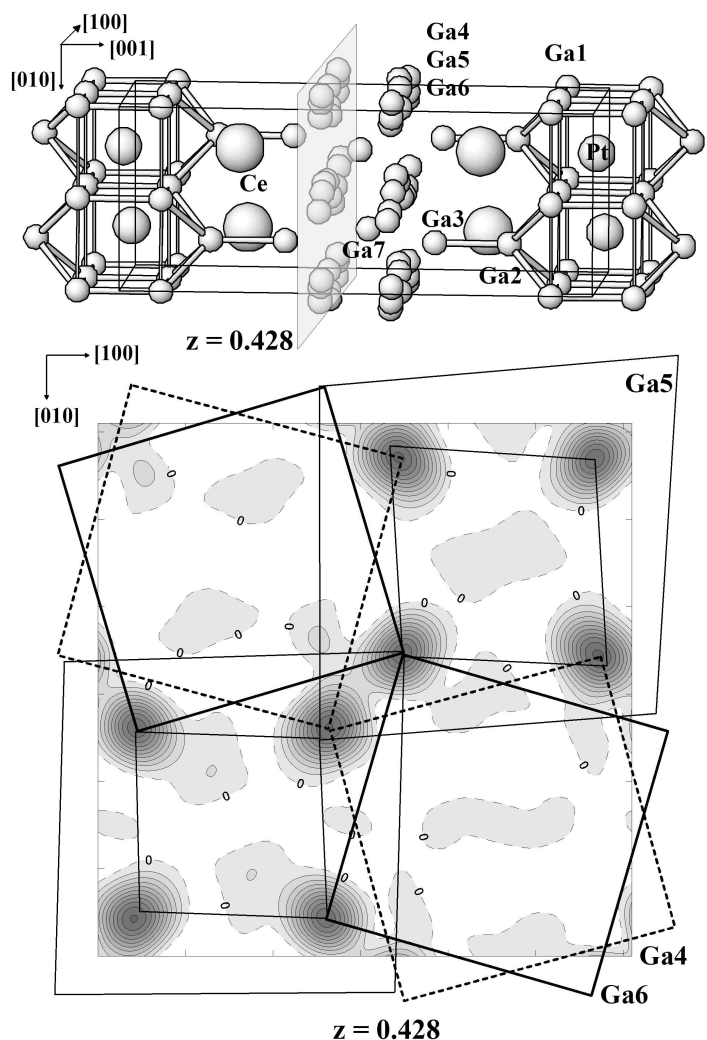}
\caption{\label{fig01} Crystal structure of \CPG\, (top) and distribution of the difference electron density in the plane at $z = 0.428$ (bottom). Dashed line shows positions of the Ga4 atoms, filled thick line depicts the Ga6 positions, and filled thin line visualises possible positions of Ga5 atoms.}
\end{center}
\end{figure}

\begin{figure}[h]
\begin{center}
\includegraphics[width=\textwidth]{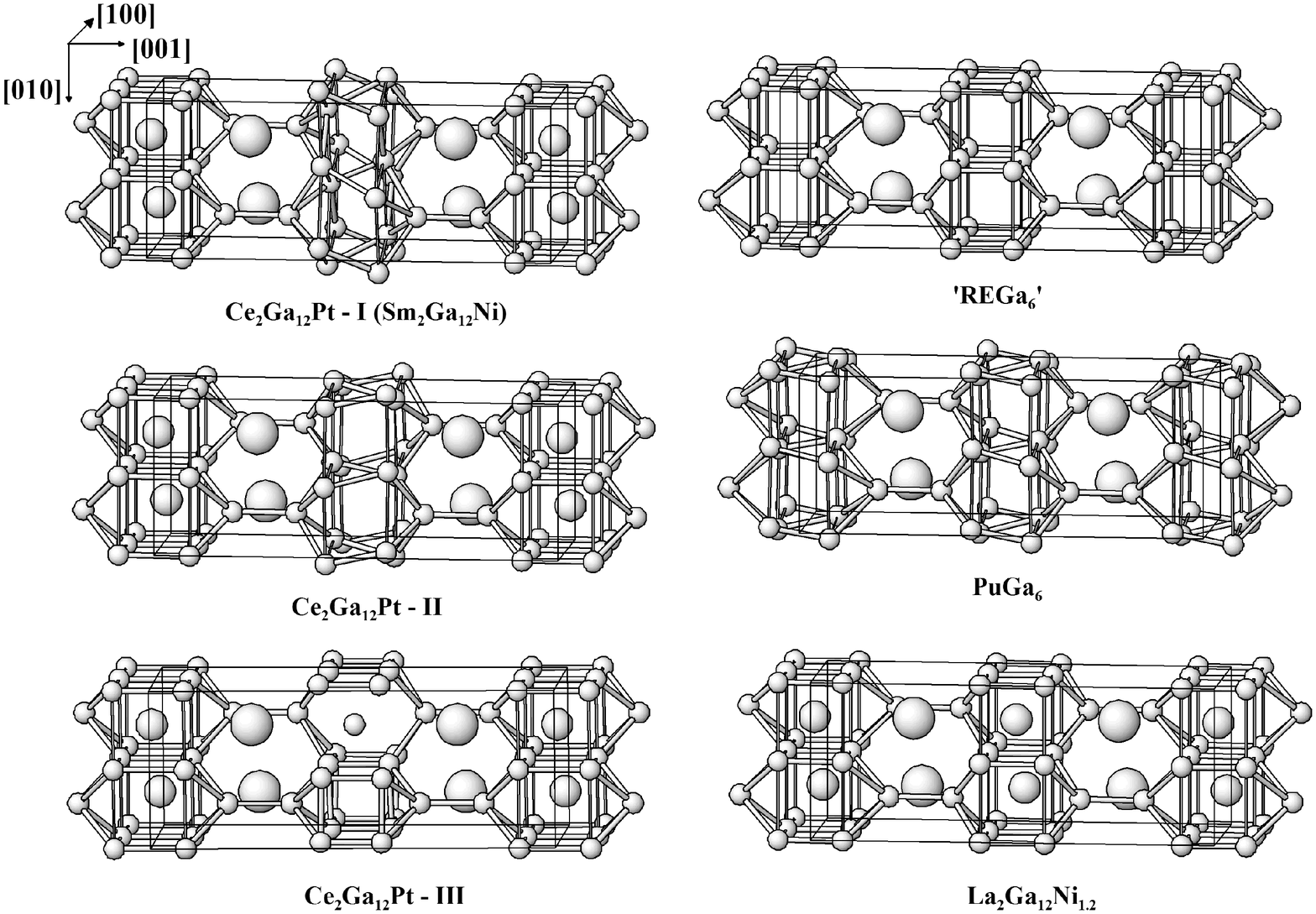}
\caption{\label{fig02} Left: Ordered models of the crystal structure of \CPG\, based on the Ga4 (Sm$_2$Ga$_{12}$Ni type, top), Ga6 (Sm$_2$Ga$_{12}$Ni type, middle) and Ga5, Ga6, Ga7 positions (bottom). Right: Hypothetical structure '$RE$Ga$_6$' with cubic cavities in the Ga framework (top); atomic arrangement in the hexagallides of $RE$ metals with the PuGa$_6$-type crystal structure (middle) and filled cubic voids in the crystal structure of La$_2$Ga$_{12}$Ni$_{1.2}$ (bottom).}
\end{center}
\end{figure}

Taking into account the lattice parameters (table \ref{tab1}) and the observed extinctions in the diffraction intensities ($hk0$ observed only with $h + k = 2n$, $0kl$ observed only with $k = 2n$) the centrosymmetric tetragonal space group $P4/nbm$ was assumed, being in agreement with the crystal structure of the Sm$_2$Ga$_{12}$Ni type \cite{x1} also found for the palladium analogous compound Ce$_2$Ga$_{12}$Pd \cite{Macaluso:2005}. Refinement of the crystal structure resulted in the low residual of $R(F) = 0.028$. Analysis of the displacement parameters at this stage revealed, that the $B_{eq}$ value for the Ga4 position is twice a large as that for other Ga-occupied sites. Moreover, this position revealed a strong anisotropy of displacement ($B_{11} = B_{22} >> B_{33}$). To clarify this issue, the difference density map at $z = 0.428$ was calculated without taking into account the Ga4 position (figure \ref{fig01}, bottom). Beside the non-spherical distribution around the Ga4 position, the map revealed additional maxima of density located symmetrically to the Ga4 position with respect of the diagonal mirror plane. Further calculation of the difference density revealed a next maximum at \footnotesize $^1/_4\;^1/_4\;^1/_2$\normalsize. In total to describe the electron density in the region $0.4 < z < 0.6$, four sites with different occupancy are necessary (figure \ref{fig01}, top; table \ref{tab2}) resulting in a reduction of the residual to $R(F) = 0.024$. Thereby, the atomic displacement parameters for all Ga positions are of the same order (table \ref{tab2}), only Ga3 shows slightly enlarged values.  The latter feature can be understood by considering ordering models of the crystal structure of \CPG. While the occupation of the Ga4, Ga5 and Ga6 positions by Ga is rather sure due to their place within the Ga network, the position Ga7 may be also occupied with platinum because of its rather cubic environment. Nevertheless the distances to the next neighbours of 2.79 \AA\, are more in agreement with the Ga1-Ga2 distances (which are not affected by crystallographic disorder) being in the range $2.63 - 2.75$ \AA\, in this structure compared with the observed Pt-Ga distance of 2.56 \AA. Thus, we assume the occupation of the Ga7 position by gallium. Finally, the total content of the unit cell Ce$_4$Ga$_{23.9(2)}$Pt$_2$ is well in agreement (within one e.s.d) with the ideal composition Ce$_2$Ga$_{12}$Pt.   

The observed partially occupied Ga site can be described using three different ordered models (figure \ref{fig02}, left). In the models I and II the position at \footnotesize $^1/_4\;^1/_4\;^1/_2$\normalsize is empty. They both represent the structural motif of the Sm$_2$Ga$_{12}$Ni type, and are enantiomorph considering the Ga4 and Ga6 sites, which are symmetrical in respect of the mirror plane running through the base diagonal parallel to $[001]$.  The model III is necessary to understand the position Ga7 at \footnotesize $^1/_4\;^1/_4\;^1/_2$\normalsize. In this case the sites Ga5, Ga6 and Ga7 have to be used. Because of the low occupation of the Ga7 site, the positions of the neighbouring sites are difficult to establish among the overlapping maxima of difference electron density, thus the probability of this atomic arrangement in \CPG\, is much lower as that of both previous. The enlarged displacement parameters for the Ga3 position are caused by the non-equivalent location of Ga3 in the ordered models of \CPG. 

The genesis of the crystal structure of \CPG\, and the observed crystallographic disorder may by understood starting with the hypothetical atomic arrangement with the composition $RE_2$Ga$_{12}$ containing segments of the BaAl$_4$-type structure (composition $RE_2$Ga$_8$, $RE$ – rare earth metal) separated by slabs of empty cubes formed by Ga atoms only (composition Ga$_4$). The total composition adds to $RE_2$Ga$_{12}$. The Ga network in such arrangement bears two types of the cavities: the large one has 18 vertices and is occupied by the $RE$ atoms, the small one has the shape of a slightly distorted cube and is empty (figure \ref{fig02}, top right). The empty cube formed by gallium atoms seems to be instable, and in the real structure of the $RE$Ga$_6$ compounds (PuGa$_6$ type \cite{x2,x3,x4} it is 'screwed' toward a tetragonal antiprism (figure \ref{fig02}, middle right). This kind of local atomic arrangement of Ga is indeed observed in \CPG\, and other representatives of the Sm$_2$Ga$_{12}$Ni type around $z = 0.5$. Another way to stabilize the cubic voids is to fill up them with an appropriate species. In this case, the local arrangement will be in agreement with the CsCl type of structure which is characteristic for the equiatomic compounds of transition metals ($TM$) with gallium and aluminum. Thus the transition metals should be also suitable for fixing of the cubic environment also in the multi-component structures. For the $RE_2$Ga$_{12}TM_x$ compounds it was first find as a substructure of La$_2$Ga$_{12}$Ni$_{1.2}$ (around $z = 0.0$ and $z = 0.5$, figure \ref{fig02}, bottom right) \cite{Grin:1982} and further for the representatives of Sm$_2$Ga$_{12}$Ni type (around $z = 0.0$) \cite{Macaluso:2005, Cho:2008}. 

The disorder in the region $0.4 < z < 0.6$ seems to be characteristic for the representatives of the Sm$_2$Ga$_{12}$Ni type. The published values of the displacement parameters \cite{Macaluso:2005, Cho:2008} are very similar to that for \CPG\, without accounting for the disorder. 

\subsection{Physical properties}
\begin{figure}[t]
\begin{center}
\includegraphics[width=0.5\textwidth]{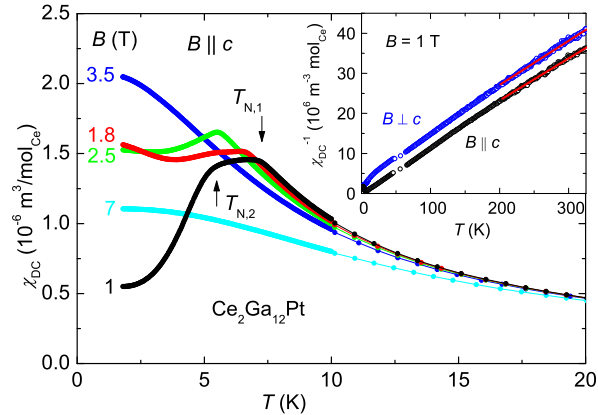}
\caption{\label{fig1}Temperature dependence of the magnetic susceptibility of \CPG\, for various magnetic fields along the $[001]$ direction. For $B = 1$\,T, the two AFM transitions are marked by arrows. In the inset the inverse susceptibility is shown for $B \parallel c$  (black symbols) and $B \perp c$ (blue symbols) together with the Curie-Weiss fits (red line).}
\end{center}
\end{figure}

We now turn to the physical characterization of a \CPG\, single crystal. The inverse susceptibility shown in the inset of figure~\ref{fig1}, with the magnetic field along and perpendicular to the $c$ direction, presents Curie-Weiss behaviour due to $4f$-derived magnetic moments. A Curie-Weiss fit above 200\,K yields an effective moment $\mu_{eff}^c=2.46\,\mu_B$ for the magnetic field along $c$ and $\mu_{eff}^{ab}=2.41\,\mu_B$ perpendicular to $c$, close to that of the free Ce$^{3+}$ moment ($2.54\,\mu_B$). The anisotropy at high temperatures is reflected in different Weiss temperatures, $\Theta_W^c=-20$\,K and $\Theta_W^{ab}=-50$\,K, for the field parallel and perpendicular to $c$, respectively. These observations can be  understood in terms of the crystal-electric-field (CEF) effect, which leads to a pronounced single-ion anisotropy with Ising-type behaviour, i.e., the magnetic easy direction is the $c$ axis of the tetragonal unit cell. However, on the basis of the present results, a complete determination of the CEF parameters cannot be given. This would need further experimental input.

In the main part of figure~\ref{fig1}, we present the temperature dependence of the \textit{dc} susceptibility, $\chi(T)$, at various magnetic fields. At $B = 1$\,T (black curve), \CPG\, undergoes two AFM transitions at $T_{N,1} = 7.3$\,K and $T_{N,2} = 5.5$\,K (see arrows in figure~\ref{fig1}). The AFM nature of these magnetic transitions is evident from the susceptibility, because in the paramagnetic phase  $\chi(T)$ is nearly field independent and drops below $T_{N,2}$. At $B = 1.8$\,T (red curve), $T_{N,2}$ is shifted to below 2\,K, whereas $T_{N,1}$ is only shifted slightly. For magnetic fields $B\geq 3.5$\,T, no transition could be resolved above 2\,K. The second transition at $T_{N,2}$ is most likely due to spin reorientation, presumably also causing a change of the ordering vector, which is commonly observed in magnetic rare earth compounds. 

\begin{figure}[t]
\begin{center}
\includegraphics[width=0.5\textwidth]{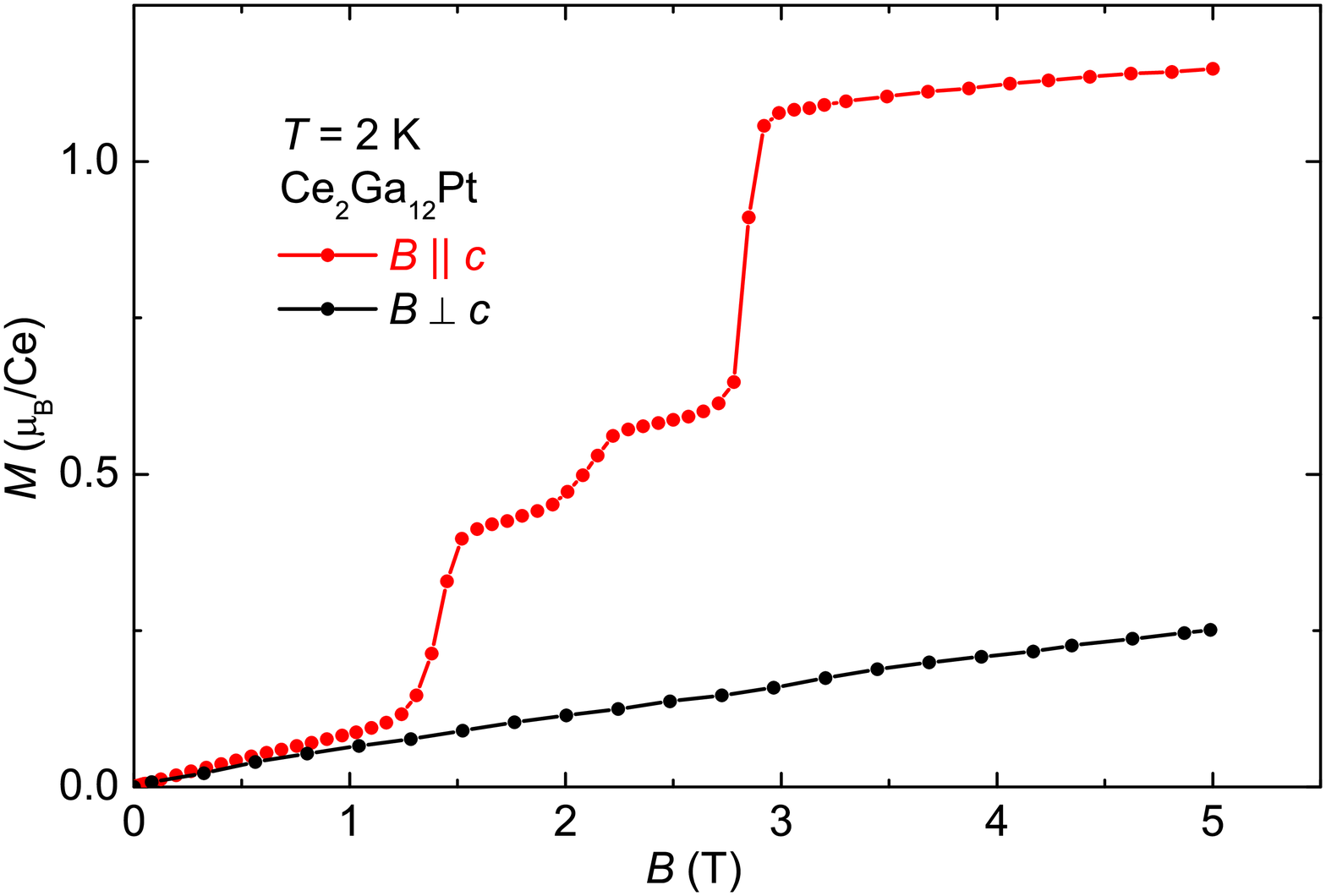}
\caption{\label{fig2}Magnetization vs. magnetic field of \CPG\, at $T = 2$\,K for the two crystallographic directions, $B \parallel c$ (red curve) and $B \perp c$ (black curve).}
\end{center}
\end{figure}

The magnetic anisotropy of \CPG\, is further apparent in the magnetization data at $T = 2$\,K, presented in figure~\ref{fig2} and measured for magnetic fields along the $c$ direction (red symbols) and in the basal-plane (black symbols), respectively. Below $B=1$\,T, both curves nearly coincidence, reflecting a rather isotropic AFM state. However, for $B\parallel  c$ three metamagnetic transitions could be resolved at 1.4, 2.1, and 2.8\,T, respectively. By contrast, for $B\perp c$ the magnetization increases almost linearly without any metamagnetic transition up to 5\,T. The magnetic anisotropy at 5\,T, $M_c/M_{ab}= 4.6$, is therefore much larger than at small fields. Presently, the precise origin of these metamagnetic transitions cannot be resolved and more microscopic probes as, e.g., neutron diffraction, are necessary to unravel the change of the magnetic structure. The comparison with magnetization data in the literature on related compounds suggests, that metamagnetic transitions along the $c$ direction are rather common in this type of compounds, as they were also observed in Ce$_2$Ga$_{12}$Pd \cite{Macaluso:2005} and in Ce$_2$Ga$_{12}$Ni \cite{Kaczorowski:2002, Cho:2008}. Whereas in \CPG\, the saturated moment along the $c$ direction ($M_{5\,T,c} = 1.15\,\mu_B$/Ce) is comparable to Ce$_2$Ga$_{12}$Ni ($M_{9\,T,c} = 1.2\,\mu_B$/Ce), it is much smaller than in Ce$_2$Ga$_{12}$Pd ($M_{5\,T,c} = 1.75\,\mu_B$/Ce).

\begin{figure}[t]
\begin{center}
\includegraphics[width=0.5\textwidth]{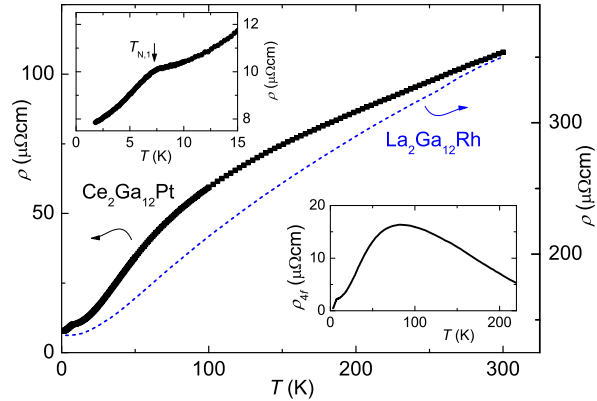}
\caption{\label{fig3}Temperature dependence of the resistivity, measured along the basal ($ab)$ plane of a \CPG\, single crystal (black symbols). Additionally, the resistivity of a nonmagnetic reference sample La$_2$Ga$_{12}$Rh is shown (dashed line) with the scale on the right hand axis. In the upper left inset the low temperature behaviour is enlarged, showing a distinct anomaly at $T_{N,1}$ (see arrow). The $4f$ contribution to the resistivity is plotted in the lower right inset (see text).}
\end{center}
\end{figure}

The electrical resistivity measured within the basal plane plotted in the main panel of figure~\ref{fig3} shows a temperature dependence typical for intermetallic Ce compounds. Three different temperature regions can be identified: (i) Above 150\,K, the resistivity of the single crystal depends nearly linearly on temperature due to phonons as discussed below. (ii) Below 150\,K, there is a deviation from the linear behaviour, probably due to reduced spin-disorder scattering at the CEF levels getting depopulated upon cooling. (iii) The first AFM transition is clearly resolved as a distinct anomaly at $T_{N,1}$. However, at $T_{N,2}$ no further anomaly was observed (see upper left inset of figure~\ref{fig3}). The overall decrease results in a resistivity  ratio, $\rho_{300\,K}/\rho_0 = 14$, reflecting the good sample quality of our single crystal, much better than that of recently reported Ce$_2$Ga$_{12}$Ni single crystals, where $\rho_{300\,K}/\rho_0 = 2.3$ \cite{Cho:2008}. However, the residual resistivity, $\rho_0=7.7\,\mu\Omega$cm is still enhanced compared to other intermetallics, probably due to the structural disorder discussed above.

To analyze the $4f$ contribution to the resistivity in \CPG\, we have synthesized a related nonmagnetic reference compound, La$_2$Ga$_{12}$Rh, because the synthesis of La$_2$Ga$_{12}$Pt was not successful. The structural  details of La$_2$Ga$_{12}$Rh will be discussed in a forthcoming publication. The susceptibility of La$_2$Ga$_{12}$Rh presents weak Pauli paramagnetism without any magnetic or superconducting transition down to 2\,K (not shown). In figure~\ref{fig3} we also present the resistivity of La$_2$Ga$_{12}$Rh with the absolute values plotted on the right hand scale. The temperature dependence is linear above 50\,K, reflecting the phonon scattering of an nonmagnetic metal. The sample quality is not as good as for \CPG\, resulting in higher absolute values of the residual resistivity, $\rho_{0}=138\,\mu\Omega$cm and a correspondingly low resistivity ratio, $\rho_{300\,K}/\rho_{0} = 2.54$. We now assume that the temperature dependence of the resistivity in \CPG\, can be described as a sum of the residual resistivity, $\rho_0$, the phonon contribution, $\rho_{ph}(T)$, and the magnetic contribution, $\rho_{4f}(T)$, i.e.  $\rho^{Ce}(T)=\rho_0^{Ce}+\rho_{ph}(T)+\rho_{4f}(T)$. Accordingly, the resistivity of La$_2$Ga$_{12}$Rh is a sum of the residual resistivity and the phonon contribution only, $\rho^{La}(T)=\rho_0^{La}+\rho_{ph}(T)$. In first approximation, $\rho_{ph}$ of \CPG\, and La$_2$Ga$_{12}$Rh should be equal, as the Debye temperatures of \CPG\, and La$_2$Ga$_{12}$Pd are rather similar (see below). Therefore, we get $\rho_{4f}(T)=\rho^{Ce}(T)-\rho_0^{Ce}-\rho_{ph}(T)$ with $\rho_{ph}=[\rho^{La}(T)-\rho_0^{La}]\times \frac{\Delta \rho^{Ce}}{\Delta \rho^{La}}$ and $\Delta \rho = \rho(300\,$K)$-\rho_0$. The last factor in the equation for $\rho_{ph}$ is just a temperature independent constant and accounts for the different sample quality of \CPG\, and La$_2$Ga$_{12}$Rh.  
In the lower right inset of figure~\ref{fig3} we have plotted the $4f$ contribution as a function of temperature. A broad maximum is visible at $T_{max}\approx  85$\,K typical for a Kondo lattice system, originating from Kondo scattering off the ground state and the excited CEF levels.

\begin{figure}[t]
\begin{center}
\includegraphics[width=0.5\textwidth]{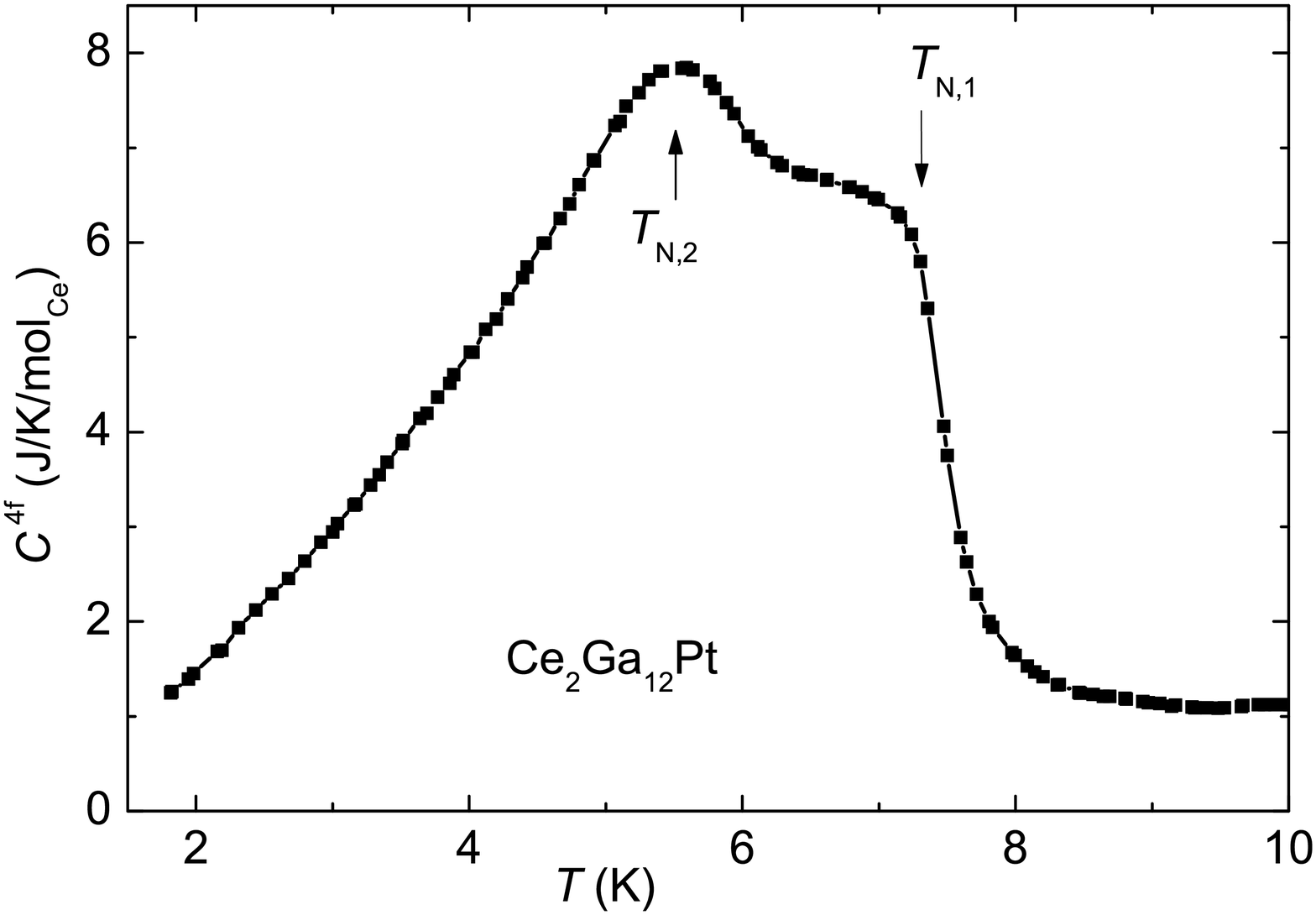}
\caption{\label{fig4}$4f$ increment to the specific heat of \CPG, calculated by subtracting the phonon contribution estimated from the literature values of the nonmagnetic La$_2$Ga$_{12}$Pd \cite{Macaluso:2005}. The two AFM transitions are marked by arrows.}
\end{center}
\end{figure}

In figure~\ref{fig4}, we show the specific heat of the same single crystal on which the magnetic measurements (see figures~\ref{fig1} and \ref{fig2}) were performed. The $4f$ contribution to the specific heat, $C^{4f}$(T), of \CPG\, was obtained by subtracting the phonon contribution, which was estimated from the literature values of the nonmagnetic La$_2$Ga$_{12}$Pd \cite{Macaluso:2005}, yielding $C^{ph}(T)=4\,$mJ$/($K$^2$mol$_{La})\times T + 1.8$mJ$/($K$^2$mol$_{La})\times T^3$. The corresponding Debye temperature of La$_2$Ga$_{12}$Pd amounts to 200\,K. Additionally, we have cross-checked the phonon contribution of \CPG\, from a linear fit of $C(T)/T$ vs. $T^2$ between 10 and 20\,K which gives a Debye temperature of 205\,K, very similar to the one of La$_2$Ga$_{12}$Pd. The resulting $4f$ contribution to the specific heat of \CPG\, presents a large double-peak anomaly at $T_{N,1,2}$, confirming the intrinsic nature of these two AFM transitions observed in $\chi(T)$. Below $T_{N,2}$, $C^{4f}(T)$ follows a quadratic temperature dependence presumably due to two-dimensional AFM magnons which is reasonable regarding the layered nature of the crystal structure. 
Extrapolating the quadratic temperature dependence to zero temperature, one finds a Sommerfeld coefficient of order $\gamma \sim 0.2$\,J$/($K$^2$mol$_{Ce})$. The magnetic entropy gain per Ce atom, calculated by integrating $C^{4f}(T)/T$ over temperature, amounts to $1.2R\ln 2$ at 8\,K  which proves that the CEF ground state is a doublet, but indicates that the first excited level is not too far above the ground state (of order 10 to 20\,K). In addition, the fact that the whole Zeeman entropy of this doublet ground state is released at the upper AFM phase transition implicates a Kondo temperature which does not exceed $T_{N,1}$.

\section{Conclusions}

We succeeded to grow single crystals of a new ternary compound \CPG, which presents an interesting tetragonal crystal structure with two-dimensional Ce planes embedded into a three-dimensional Ga-Pt framework showing, on top, crystallographic disorder. The magnetic anisotropy due to the CEF effect is revealed by magnetic-susceptibility measurements and presents Ising-type behaviour with the easy magnetic axis along $c$. \CPG\, orders antiferromagnetically at $T_{N,1}=7.3$\,K followed by a second AFM transition at $T_{N,2}=5.5$\,K, most probably due to spin reorientation of the $4f$ moments. Several metamagnetic transitions are observed when the field is applied along $c$, typical for this type of compounds. Specific-heat measurements prove the intrinsic nature of the magnetic transitions and reveal a doublet CEF ground state with a related Kondo temperature $T_K\lesssim T_{N,1}$. Resistivity measurements further support classification of \CPG\, as a magnetically ordered Kondo-lattice system in the weak-coupling regime of the Doniach phase diagram \cite{Doniach:1977}.

\end{document}